\documentclass[letterpaper,11pt]{article}
\usepackage{fullpage}  

\begin{document}

\addtolength{\textheight}{+0.18in}
 
\newtheorem{lemma}{Lemma}[section]
\newtheorem{theorem}{Theorem}
\newtheorem{cor}{Corollary}[section]
\newtheorem{prop}{Proposition}[section]
\newcommand{\lf}{\left\lfloor}
\newcommand{\rf}{\right\rfloor}
\newcommand{\dis}{\displaystyle}
\newcommand{\sums}[2]{\sum_{\stackrel{{\scriptstyle {#1}}}{{#2}}}}
\newcommand{\prods}[2]{\prod_{\stackrel{{\scriptstyle {#1}}}{{#2}}}}
\newtheorem{Definition}{Definition}
\newtheorem{corollary}{{Corollary}}

\newenvironment{tabAlgorithm}[2]{
\setcounter{algorithmLine}{1} \samepage
\begin{tabbing}
999\=\kill #1 \ \ --- \ \ \parbox{3.8in}{\it #2} }{
\end{tabbing}
}
\newcounter{algorithmLine}
\newcommand{\algline}{\\\thealgorithmLine\hfil\>\stepcounter{algorithmLine}}
\newcommand{\algnono}{\\ \>}

\newcommand{\TRUE}{{\bf TRUE}}
\newcommand{\FALSE}{{\bf FALSE}}
\newcommand{\NIL}{{\bf NIL}}
\newcommand{\CURRENT}{{\bf CURRENT}}
\newcommand{\IF}{{\bf IF }\=}
\newcommand{\THEN}{{\bf THEN }\=}
\newcommand{\ELSE}{{\bf ELSE }}
\newcommand{\WHILE}{{\bf WHILE }\=}
\newcommand{\FOR}{{\bf FOR }\=}
\newcommand{\DO}{{\bf DO }\=}
\newcommand{\RETURN}{{\bf RETURN }}
\newcommand{\BREAK}{{\bf BREAK }}

\input epsf
\def\prbox{\hfill\rule{1.2ex}{1.2ex}\vspace{.2in}}
\newenvironment{proof}{\noindent{\bf Proof }}{\prbox}
\def\eps{\epsilon} 

\title{Minimum Enclosing Polytope in High Dimensions
}

\author{ 
Rina Panigrahy\thanks{ Cisco Systems, San Jose, CA 95134. E-mail: {\tt rinap@cisco.com}.}
}

\maketitle
\begin{abstract}
We study the problem of covering a given set of $n$ points in a high, $d$-dimensional space by the 
minimum enclosing polytope of a given arbitrary shape. We present algorithms that work
for a large family of shapes, provided either only translations and no rotations are
allowed, or only rotation about a fixed point is allowed; that is, one is allowed to only scale and translate
a given shape, or scale and rotate the shape around a fixed point. Our algorithms start with 
a polytope guessed to be of optimal size and iteratively moves it based on a greedy principle: simply
move the current polytope directly towards any outside point till it touches the surface.
For computing the minimum enclosing ball, this gives a simple greedy algorithm with running time 
$O(nd/\eps)$ producing a ball of radius $1+\eps$ times the optimal. 
This simple principle generalizes to arbitrary convex shape when only translations are allowed, 
requiring at most $O(1/\eps^2)$ iterations. Our algorithm implies that {\em core-sets} of size $O(1/\eps^2)$ 
exist not only for minimum enclosing ball but also for any convex shape with a fixed orientation.
A {\em Core-Set} is a small subset of $poly(1/\eps)$ points whose minimum enclosing
polytope is almost as large as that of the original points. When only rotation about a fixed point
is allowed, for a certain class of convex 
bodies with an axis of symmetry that includes cylinders, cones and ellipsoids, we prove
that our techniques work provided the problem is confined to a half space.
Without the half-space restriction, we obtain an algorithm whose running time is 
exponential in $1/\eps^2$, and corresponding core-sets. 
This automatically gives us an  $2^{O(1/\eps^2)}nd$ time algorithm for the 
min-cylinder problem provided we are given a fixed point on the axis. Although we are unable to
combine our techniques for translations and rotations for general shapes, for the min-cylinder
problem, we give an algorithm similar to the one in \cite{HV03}, but 
with an improved running time of $2^{O(\frac{1}{\eps^2}\log \frac{1}{\eps})} nd$.
This generalizes to computing 
the minimum radius $k$-dimensional flat in time $exp(\frac{e^{O(k^2)}}{\eps^2}\log \frac{1}{\eps}) nd$.
\end{abstract}

%\newpage
 
\section{Introduction}
Given a set $S$ of $n$ points in $d$ dimensions, we study the problem of finding the 
minimum enclosing polytope of a given arbitrary shape when $d$ is large. Being a fundamental
problem in computational geometry with applications
in data mining, learning, statistics and clustering (\cite{GK93}, \cite{GK94}, \cite{HV02}), this problem has 
a  rich history. B\u{a}doiu et. al. \cite{BHI02} gave an algorithm that computes the minimum enclosing 
ball approximately,  with radius at most $1+\eps$ times 
the optimal radius in time $O(nd/\eps^2 + (1/\eps)^{10})$, independent of the number of dimensions, using
convex programming. 
 Their algorithm was based on the idea of {\em Core-Sets}, a small set of 
$poly(1/\eps)$ points whose minimum enclosing ball is almost as large as that of all the $n$ points.
This was improved to $O(nd/\eps + (1/\eps)^{5})$ in 
\cite{MK03} by finding smaller core-sets of size $\lceil 1/\eps \rceil$. They also provide a simple
$O(nd/\eps^2)$ time algorithm for finding the minimum enclosing ball that does not require
convex programming. Combining the two results gives a $O(nd/\eps + (1/\eps)^{5})$ time algorithm for
finding the minimum enclosing ball while eliminating the use of convex programming.

For the minimum enclosing cylinder problem, Har-Peled and Varadarajan \cite{HV03} gave an algorithm
with running time of $2^{O(\frac{1}{\eps^3} \log^2 \frac{1}{\eps})}nd$ that finds a cylinder with 
radius at most $1+\eps$ times the optimal radius. They also generalized their algorithm to computing
the minimum radius $k$-dimensional flat in time $exp(\frac{e^{O(k^2)}}{\eps^{2k+3}}) nd$,
 where the radius of a $k$-flat is the maximum
distance of the given set of points from this $k$-flat. 

In this paper, we present algorithms for computing the minimum enclosing polytope 
for a large family of shapes, provided either only translations and no rotations are
allowed, or only rotation about a fixed point is allowed; that is, one is allowed to only 
scale and translate
a given shape, or scale and rotate the shape around a fixed point. We hope that it may be possible to
combine the techniques for translation and rotation to solve the problem without these
restrictions. Our algorithms are based on a simple greedy principle applied iteratively: simply
move the current polytope directly towards any outside point till it touches the surface.

For computing the minimum enclosing ball, this gives a simple greedy algorithm that repeatedly moves
a ball directly towards the farthest uncovered point till it touches the surface. If we start 
with a ball of the optimal radius, we show that running $O(1/\eps)$ such steps gives the 
optimal position of the ball approximately, within a running time of $O(nd/\eps)$ (section \ref{sec1}). 
This simple principle generalizes to arbitrary convex shape when only translations are allowed, 
requiring at most $O(1/\eps^2)$ iterations (section \ref{sec2}). It also works if 
the shape can be expressed as a union of a few convex shapes -- however, requiring a running time
exponential in $1/\eps^2$. Our algorithm implies that core sets of size $O(1/\eps^2)$ 
exist not only for the minimum enclosing ball but also for any convex shape with a fixed orientation.

Next we look at covering a set of points by a convex body
while allowing only rotation about a fixed point (section \ref{sec3}). For a certain class of convex 
bodies with an axis of symmetry  that includes cylinders, cones and ellipsoids, we prove
that our techniques work provided the problem is confined to a half space bordering at the 
point of rotation. Without this restriction, we obtain an algorithm whose running time is 
exponential in $1/\eps^2$. This gives us an  $2^{O(1/\eps^2)}nd$ time algorithm for the 
min-cylinder problem provided we are given a fixed point on the axis. This also
implies that core-sets whose size depend only on $\eps$ exist for rotational problems as well.
 Although we are unable to
combine our techniques for translations and rotations for general shapes, for the min-cylinder
problem, we give an algorithm almost identical to the one in \cite{HV03}, but 
with an improved running time of $2^{O(\frac{1}{\eps^2}\log \frac{1}{\eps})} nd$ (section \ref{sec4}).
This generalizes to computing 
the minimum radius $k$-dimensional flat in time $exp(\frac{e^{O(k^2)}}{\eps^2}\log \frac{1}{\eps}) nd$.

\section{Minimum Enclosing Ball}\label{sec1}
 Given a set $S$ of $n$ points in $d$ dimensions, we provide an algorithm to compute the 
minimum enclosing ball with radius at most $1+\eps$ times the optimal in time $O(nd/\eps)$.

We start with a simple algorithm $MEB$ (figure \ref{code-MEB}) that works as follows: the algorithm 
starts with an arbitrary ball of the optimal radius, and for any point at least $\eps$ 
outside this ball, moves the ball till the surface touches the outside point.
This involves guessing the optimal radius of the minimum enclosing ball. Assume without loss 
of generality that the optimal ball is of unit radius.

Let $d(P,Q)$ denote the distance between two points $P$ and $Q$. Let $B(C,r)$ denote
the ball of radius $r$ centered at point $C$.

\begin{figure}
{\centering \fbox{\begin{minipage}{\columnwidth}
\begin{tabAlgorithm}{{\bf Algorithm MEB}}{}
\algline Start with a ball of optimal radius.

\algline Repeat \= until every point is within $1+\eps$ of the current center $C$.

\algline \>Find the farthest point $P$ from $C$.

\algline \>Move $C$ towards the point $P$ till $P$ touches \\ \> \> the unit
	sphere centered at $C$.    
\end{tabAlgorithm}
\end{minipage}
} \caption{A simple algorithm for finding the minimum enclosing ball}\label{code-MEB}}
\end{figure}

\begin{figure}
\begin{center}
\epsffile[48 551 226 760]{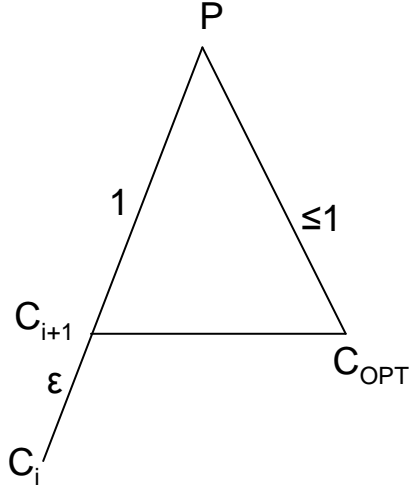}
\caption{The center of the ball gets closer to the optimal position in each iteration}\label{fig1}
\end{center}
\end{figure}

\begin{theorem}\label{thm1}
Algorithm $MEB$ terminates in $O(1/\eps)$ iterations.
\end{theorem}

The basic idea behind this theorem is that in each iteration the center $C$ 
in the algorithm moves closer to the optimal center $C_{OPT}$.
If $d_i$ is the distance of $C$ from $C_{OPT}$ in the $i^{th}$ iteration, then $d_i$
decreases as follows.

\[d_{i+1}^2 \le d_i^2 - \eps^2\]

This is because $\angle C_i, C_{i+1}, C_{OPT}$ is obtuse (figure \ref{fig1}), as
$PC_{i+1}$ is not shorter than $PC_{OPT}$ implying $\angle PC_{i+1}C_{OPT}$ is acute.
So, $d_{i+1}^2 \le d_i^2 - d(C_i, C_{i+1})^2 \le d_i^2 - \eps^2$.
Since the initial value of $d_i^2$ is at most a constant, and since it decreases by $\eps^2$ in
each iteration, the algorithm must terminate in $O(1/\eps^2)$ iterations.
A tighter analysis based on the following lemma 
will show that it actually terminates in $O(1/\eps)$ iterations.

\begin{lemma}\label{goodh}
If $C$ and $C_{OPT}$ are distance $d$ apart, there must be a point that is at least
$d^2/4$ from the surface of the unit ball centered at $C$.
\end{lemma}
\begin{proof}
Look at the hemisphere in the optimal ball that is directly facing away from $C$. It is well known
(for reference see lemma 5.2
in \cite{GIV01}) that one of the $n$ points, $P$, must lie on this hemisphere. Since $\angle C C_{OPT} P$
is obtuse,
\[d(C, P) \ge \sqrt{1+d^2}
	   \ge 1+d^2/4 \mbox{ ( since } d \le 2)\]

This means that the point $P$ is at least $d^2/4$ away from the surface of the unit ball 
centered at $C$. 
\end{proof}

Now that we have a lower bound of $d^2/4$ on $d(C_i, C_{i+1})$, we are 
ready to prove that algorithm $MEB$ terminates in $O(1/\eps)$ iterations.

\begin{proof}{\bf of theorem \ref{thm1}.}
Note that 
$d_{i+1}^2 \le d_i^2 - d(C_i, C_{i+1})^2 
		\le d_i^2 - d_i^4/16$

Let $\Phi_i = d_i^2$, and we get the recurrence relation $\Phi_{i+1} = \Phi_i - \Phi_i^2/16$. 
It is easy to check that if we start with $\Phi_0=1$, then after $O(2^i)$ iterations,
$\Phi_i$ decreases to $\le 1/2^i$ (if $\Phi$ is $2^{-i}$, in $O(2^i)$ iterations it will 
become less than $2^{-(i+1)}$ ).  So after $O(1/\eps)$ iteration, $\Phi_i$ becomes $\le \eps$. 
After that, since it decreases by at least $\eps^2$ in each iteration, there can be at most
$\eps/\eps^2 = 1/\eps$ further iterations.
\end{proof}

Algorithm $MEB$ requires guessing the optimal radius. The distance of the farthest point
from any given point in the set is within factor $2$ of the optimal radius. 
 A binary search with at most
$O(\log(1/\eps))$ tries can be used to ascertain the correct radius approximately
within a factor of $1+\eps$. If a guess is 
too small the algorithm will not terminate in $O(1/\eps)$ iterations.
For a certain guess, if the algorithm terminates
successfully, this means the guess is greater than 
a $1+\eps$ approximation of the optimal radius and so the guess may be decreased. 
So by running $MEB$ at most $O(\log(1/\eps))$ times
the minimum enclosing ball can be computed in time $O(\frac{dn}{\eps}\log\frac{1}{\eps})$.

\subsection{Eliminating the Binary Search}

To eliminate the binary search, in algorithm $MEBOPT$ (figure \ref{code-MEBOPT}), 
we start  with a ball of radius less than optimal
 and increase it in certain iterations.
The radius of the ball is always less than the optimal radius but
the gap decreases as the iterations proceed.

\begin{figure}
{\centering \fbox{\begin{minipage}{\columnwidth}
\begin{tabAlgorithm}{{\bf Algorithm MEBOPT}}{}

\algline Initi\=alize \=  $C =$ any arbitrary point of the $n$ points

\algline \>		$r = 1/2$ of distance of farthest point from any one point

\algline \>		$\delta = 1/2$ of distance of farthest point from any one point

\algline Repeat until $\delta \le \eps$

\algline \>	For $O(1/\delta)$ iterations

\algline \>\>		find farthest point $P$ from $C$ 
\algline \>\>		move $B(C,r)$ till its surface touches $P$.

\algline \>	Let $s$ = distance of the farthest point from the surface of current ball $B$

\algline \>	If $s \le 3\delta/4$

\algline \>\>		$\delta = 3\delta/4$

\algline \>	else $r = r + \delta/4$

\algline \>\>		$\delta = 3\delta/4$
  
\end{tabAlgorithm}
\end{minipage}
} \caption{An algorithm that does not require binary search}\label{code-MEBOPT}}
\end{figure}

The algorithm maintains a lower bound $r$, and an error $\delta$, such that 
$r \le r_{OPT} \le r+\delta$. Again as before we assume that the optimal radius is $1$. 
In each iteration we reduce the error $\delta$ by a factor of $3/4$ based on the following lemma.

\begin{lemma}
After $O(1/\delta)$ iterations of moving the ball to farthest outside point, every outside 
point must be within $3\delta$ from the surface.
\end{lemma}
\begin{proof}
Let $h_i$ denote the distance of farthest point from surface of $B(C_i, r_{OPT})$, 
after $i$ iterations. Then the farthest point is at a distance $h_i - \delta$ from
$B(C_i, r_{OPT})$. Just as in proof of theorem \ref{thm1},  
it is easy to check that (only difference is that $d(C_i, C_{i+1}) = h_i - \delta$ )

\[d_{i+1}^2 \le d_i^2 - (h_i - \delta)^2\]

Now as long as $h_i \ge 2\delta$, we have $d_{i+1}^2 \le d_i^2 - h_i^2/4$. Again
lemma \ref{goodh} says $h_i \ge d_i^2/4$, implying that as long as 
$h_i \ge 2\delta$, we have $d_{i+1}^2 \le d_i^2 - d_i^4/64$
So as in proof of theorem \ref{thm1}, in $O(1/\delta)$ iterations, either $d_i \le \delta$ or 
$h_i \le 2\delta$. In either case, distance of the farthest point from the center
is at most $r_{OPT} + 2\delta \le r + 3\delta$.

\end{proof}

Using the above lemma, we can test if the current estimate $r$ in fact has an error of at most
$\delta/4$ in $O(4/\delta)$ iterations. If after so many iterations, the farthest point 
is more than $3\delta/4$ away from the surface, then we can conclude that the error was more 
than $\delta/4$ and so $r$ can be increased to $r+\delta/4$. Otherwise, since every point 
is within distance $3\delta/4$ outside the surface, we can conclude
that the error in $r$ is at most $3\delta/4$. Since each iteration in step 4 runs in
time $O(nd/\delta)$, and $\delta$ decreases geometrically to $\eps$,
we have proved the following theorem. 

\begin{theorem}\label{thm2}
Algorithm $MEBOPT$ finds an approximate minimum enclosing ball in time $O(nd/\eps)$
\end{theorem}

\section{Generalizing to Convex Polytopes}\label{sec2}

The simple algorithm of moving towards the outside point works not only for finding the
minimum enclosing ball but also for minimum enclosing polytope of any given convex shape
with a fixed orientation. That is, one is only allowed to translate and scale the given
shape but is not allowed to rotate it. Again, for ease of exposition, we assume that
the maximum inter-point distance is at most $1$. We present algorithm $MINCON$ (figure \ref{code-MINCON}),
similar to $MEB$, that finds an approximate
optimal solution in $1/\eps^2$ iterations. 
Again as in algorithm $MEB$, we guess the optimal size of the given
shape but do not know its position to begin with. We repeatedly find an outside point
at least $\eps$ away from the surface and move the current polytope by the shortest 
distance till the point touches the surface.

\begin{figure}
{\centering \fbox{\begin{minipage}{\columnwidth}
\begin{tabAlgorithm}{{\bf Algorithm MINCON}}{}

\algline  Start with a polytope guessed to be of optimal size positioned anywhere.

\algline  Repeat \= until done

\algline \>Find any point $P$ that is at least $\eps$ away from the \\
	\>\>surface of the current polytope

\algline \>Find the point $Q$ on the polytope that is closest to $P$.

\algline \>Move the polytope so that $Q$ coincides with $P$.
  
\end{tabAlgorithm}
\end{minipage}
} \caption{Algorithm for finding Minimum Enclosing Convex Polytope}\label{code-MINCON}}
\end{figure}

\begin{figure}
\begin{center}
\epsffile[60 585 204 750]{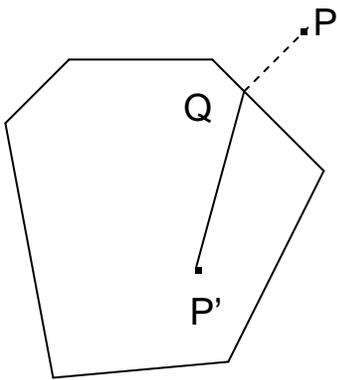}
\caption{The polytope keeps moving closer to the optimal position in each iteration}\label{fig2}
\end{center}
\end{figure}

\begin{theorem}\label{thm4}
Algorithm $MINCON$ terminates in $1/\eps^2$ iterations. That is, after so many iterations
no point will be more than $\eps$ outside the surface.
\end{theorem}

\begin{proof}
The proof is very similar to that of theorem 1. Let $Q$ be the point on the current polytope
closest to $P$ (figure \ref{fig2}). Also, we know that $P$ is in the optimal polytope. Let $P'$ denote the
corresponding point in the current polytope. That is, the vector $\overline{P' P}$ is the displacement
of the optimal polytope from the current polytope. We will prove that $\angle P' Q P$ is obtuse.
If not, there is a point on the segment $P' Q$ that is closer to $P$ than $Q$. And since both $P'$ and
$Q$ are in the current polytope, that point must also be within the current polytope. So $Q$ cannot be the
closest point to $P$, which is a contradiction. If $d_i$ denotes the distance 
of the current polytope of the optimal one in this $ith$ iteration, then 
$d_i = d(P', P)$. After the displacement by the vector $\overline{Q P}$, the polytope 
will be off from the optimal position by the vector $\overline{P' P} - \overline{Q P} = \overline{P' Q}$.
So $d_{i+1}^2 = d(P',Q)^2 \le d(P',P)^2 - d(Q, P)^2 \le d_i^2 - \eps^2$
This proves that the algorithm terminates in $1/\eps^2$ iterations. 
\end{proof}

This technique also extends to shapes that are a union of a small number of convex-shapes

\begin{theorem}\label{thm-un-con}
Given a shape and orientation that can be expressed as a union of $c$ convex-shapes,
 the smallest
enclosing polytope with that shape and orientation can be computed within $\eps$ approximation
 in time $c^{O(1/\eps^2)}nd$
\end{theorem}

\begin{proof}
The algorithm is identical to  $MINCON$, except that at each step we guess one of the convex bodies that
contains the outside point and move the polytope till that convex body touches the point. 
\end{proof}

This in fact proves that Core-Sets exist not only for minimum enclosing ball but also for 
any convex shape with a given orientation. 
\begin{Definition}
Given a set of points, $S$, and convex shape and orientation, we say that a subset $T$
of $S$ forms a Core-Set if the minimum enclosing 
polytope of $T$ has every point of $S$ within 
distance at most $\eps$ outside its surface. 
\end{Definition}

\begin{theorem}
For a set of points  with maximum inter-point distance $1$, and for a given convex 
shape and orientation, there is a Core-Set of size $O(1/\eps^2)$.
\end{theorem}

\begin{proof}
Instead of starting with a polytope of the optimal size, we start with one just small enough 
so that it can never be positioned to have every outside point within distance $\eps$ from its 
surface. Now we know that if we start with this size, then algorithm $MINCON$ would never terminate
in the $O(1/\eps^2)$ iterations it otherwise would have. We let the algorithm run for one more 
than $O(1/\eps^2)$ iterations and look at the $1+O(1/\eps^2)$ points that are visited. We will
prove that these points form the required Core-Set. The 
minimum enclosing polytope of these $1+O(1/\eps^2)$ points must be larger than the one we started 
with as otherwise, by theorem \ref{thm4}, algorithm $MINCON$ would not require more than $1+O(1/\eps^2)$ iterations
on these points. Since the initial polytope can be chosen so that every outside point is within
distance arbitrarily close to $\eps$ from the surface of this initial polytope, we have proved the 
theorem. 
\end{proof}

\section{Allowing Rotations}\label{sec3}

So far we did not allow the convex polytope to be rotated and only allowed translations. 
In this section we prove that our 
techniques work if only rotation about a fixed point and no translations are allowed, provided
 certain conditions are met.

Given a polytope that has an axis of symmetry (that is, every cross section along the axis is 
hyper-sphere of dimension $d-1$) with the axis passing through the origin, and a set of points $S$, our goal is to rotate
the axis till the points in $S$, are covered by the polytope. We will also assume that the following
conditions are satisfied

\begin{itemize}

\item All these points and the optimal polytope lie in a half-space with the bounding hyper-plane 
passing through the 
origin.

\item Any $d$-dimensional hyper-sphere centered at the origin intersects the polytope in a single
hyper-sphere of dimension $d-1$. This $d-1$-dimensional hyper-sphere divides the original
$d$-dimensional hyper-sphere into two disjoint regions.  We also assume
that the interior of the polytope intersects
the $d$-dimensional hyper-sphere in the smaller of these two regions.
(this is similar to the convexity requirement in algorithm $MINCON$. 
In three dimensions this would mean that every sphere passing through the origin
cuts the polytope in at most one circle.  Also the interior of the polytope intersects
the sphere in the smaller of the two regions on the sphere formed by the circle).

\end{itemize}

 Examples that satisfy these conditions are
cylinders, half-cones, ellipsoids lying in a half-space with axis passing through the origin. Again, for 
ease of exposition, we assume that all points are at most at unit distance from the origin.
Our algorithm $MINROT$ (figure \ref{code-MINROT}) repeatedly rotates the polytope by the smallest
angle so as to touch an uncovered point at least $\eps$ outside the surface. 

\begin{figure}
{\centering \fbox{\begin{minipage}{\columnwidth}
\begin{tabAlgorithm}{{\bf Algorithm MINROT}}{}

\algline  Start with the axis as any ray in the given half-space shooting from the origin.

\algline Iteratively find any outside point and rotate the axis by the smallest\\ 
	\> angle till the surface of the polytope touches the outside point. \\
	\>We assume that the distance between the outside point and the point  \\
	\>on the surface  it touches is at least $\eps$, as otherwise we are done.

\end{tabAlgorithm}
\end{minipage}
} \caption{Algorithm for rotational problem in a half-space}\label{code-MINROT}}
\end{figure}

\begin{figure}
\begin{center}
\epsffile[64 536 322 750]{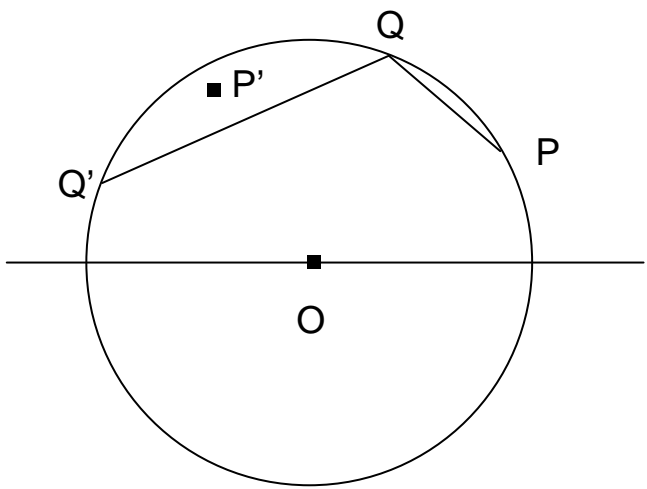}
\caption{The axis keeps rotating to the optimal position in each iteration}\label{fig3}
\end{center}
\end{figure}

\begin{theorem}\label{thm5}
Algorithm $MINROT$ terminates in $1/\eps^2$ iterations. That is, after so many iterations
no point will be more than $\eps$ outside the surface. 
\end{theorem}

\begin{proof}
The proof is very similar to that of theorem \ref{thm1}.
Let $\theta$ be the angle between the current axis and the optimal one. We will argue that
the distance between corresponding points on the two axes on the units sphere centered at the	
origin decreases in each iteration. This distance $d = 2\sin(\theta/2)$.

Let $P$ be the outside point chosen in a certain iteration and $Q$ be the closest point on 
the current polytope in terms of rotation required to move $Q$ to $P$. Look at the sphere
centered at the origin passing through $P$ (figure \ref{fig3}). This sphere intersects the 
current polytope in
a hyper-circle, $C$, passing through $Q$. The point $P$ lies within the optimal polytope. Let $P'$
be the corresponding point in the current polytope. $P'$ must be on the sphere inside the 
hyper-circular region $C$.
 Look at the great circle on the sphere passing through $P$
and $Q$. Project all points onto the two dimensional space containing this great circle.
Under this projection the hyper-circle $C$ will become a segment $QQ'$. Since the points $P, Q, Q'$ 
lie on a half circle $\angle P Q Q'$ is obtuse. Since, under the projection, $P'$ lies in 
the minor segment formed by $QQ'$, $\angle P Q P'$ is also obtuse. 
Scale the distances so that the great circle is of unit radius. 
So $d_{i+1}^2 = d(P',Q)^2 \le d(P',P)^2 - d(Q,P)^2 \le d_i^2 - \eps^2$ 
\end{proof}

Algorithm MINROT assumes that the polytope lies in a given half-space. We now provide an alternate
algorithm for polytopes with an axis of symmetry passing through the origin without the half-space
assumptions in MINROT. We will assume that the polytope is symmetric around the origin and its 
intersection with any hyper-sphere is at most two equal sized disjoint $d-1$-dimensional 
hyper-spheres on opposite sides of the origin.  Again the interior of the polytope
intersects the original hyper-sphere in the smaller of the two regions formed by each of the two
$d-1$-dimensional intersection hyper-spheres.
Examples are cylinders, cones, ellipsoids centered at the origin.
However this algorithm runs in time $2^{O(1/\eps^2)}nd$.  For ease of exposition,
this algorithm, $FULLROT$ (figure \ref{code-FULLROT}), is described for three dimensions.

\begin{figure}
{\centering \fbox{\begin{minipage}{\columnwidth}
\begin{tabAlgorithm}{{\bf Algorithm FULLROT}}{}

\algline  Start with the axis as any ray from the origin.

\algline  For each outside point $P$ look at the sphere centered at the origin.

\algline The sphere intersects
the polytope in at most two circles,\\ \> $C_1$ and $C_2$ on different sides of the origin.

\algline Now the
axis could be rotated to either touch $C_1$ or $C_2$ to $P$.\\ \> Guess one of them and rotate the
axis by the smallest \\ \> angle till $P$ touches the chosen circle. 

\end{tabAlgorithm}
\end{minipage}
} \caption{Algorithm without the half-space assumption}\label{code-FULLROT}}
\end{figure}

\begin{theorem}
Algorithm $FULLROT$ terminates in $1/\eps^2$ iterations. That is, after so many iterations
no point will be more than $\eps$ outside the surface. The deterministic version of this 
algorithm runs in time $2^{O(1/\eps^2)}nd$ by trying all possible guesses.
\end{theorem}

\begin{proof}
As in the proof of theorem 4 in any iteration there would be two points $Q_1$ and $Q_2$ on the
circles $C_1$ and $C_2$ closest to the outside point $P$. As before $P'$ would lie in one of
 the minor segment
formed by one of $C_1$ and $C_2$. We guess the correct one, say $C_1$. Again project all points
to the plane containing the great circle passing through $P$ and $Q_1$. Since the angle
$Q_2 Q_1 Q_1'$  is $90$ deg, the angle $P Q_1 Q_1'$ is obtuse. The rest of the proof is same as
that of theorem \ref{thm5}. 
\end{proof}

Again, as before, we can extend our techniques to shapes that are a union of a small number of bodies
that satisfy the conditions required by algorithm $FULLROT$.

\begin{theorem}\label{thm-un-rot}
Given a shape and orientation that can be expressed as a union of $c$ shapes, each satisfying
the conditions required by algorithm $FULLROT$, we can find the smallest
enclosing polytope with that shape and orientation in time $(2c)^{O(1/\eps^2)}nd$
\end{theorem}

Just as in section \ref{sec2}, we can derive core-sets for rotational problems.
\begin{theorem}
For rotational problems with shapes satisfying conditions for algorithms $MINROT$ and
$FULLROT$, core-sets of sizes $O(1/\eps^2)$ and $2^{O(1/\eps^2)}$ exist, respectively.
\end{theorem}

Note that an infinite cylinder with its axis passing through the origin satisfies 
the assumptions of algorithm $FULLROT$. So we have:

\begin{corollary}
For the minimum radius cylinder problem if we are given a point on the axis
of the optimal cylinder, algorithm $FULLROT$ runs in time $2^{O(1/\eps^2)}nd$
\end{corollary}

Note that in the min-cylinder problem, the maximum distance between all points may not be $1$ as 
assumed. This can be easily overcome by setting the initial position of the axis to pass through
the farthest point from the origin - we omit the details here.

\section{Minimum Radius Cylinder}\label{sec4}

Although we do not have any general results for a combination of rotation and translation
for different shapes, we 
provide an algorithm for the min-cylinder problem without restrictions 
that runs in time $2^{O(1/\eps^2)}nd$.
The algorithm is similar to the one mentioned in \cite{HV03} with a running
time of $2^{O(\frac{1}{\eps^3} \log^2 \frac{1}{\eps})}nd$. 
Our algorithm can be viewed as following the greedy principle underlying the other algorithms
of this paper: In each iteration
it moves the axis of the cylinder along the plane containing the axis and an outside point by guessing 
its optimal position in that plane approximately.

Without loss of generality assume that the optimal radius is $1$.
We will show later how this optimal radius can be computed approximately using a binary search.
We start with a certain initial position of the axis that will be specified latter.
Let $l_{OPT}$ be the axis of the optimal cylinder. Let $U$ and
$V$ be the farthest two points on $l_{OPT}$ that are projections 
of points in  set $S$ on $l_{OPT}$.\\

\noindent {\bf Algorithm MINCYN -}
\begin{enumerate}
\item We iteratively compute an estimate $l_i$ of $l_{OPT}$ and
points $U_i$ and $V_i$ on $l_i$ that are  close to projections of $U$ and
$V$  on $l_i$.

\item 
 We maintain the following invariant: $d(U_i, U) \le 5$ and $d(V_i, V) \le 5$

\item $l_{i+1}, U_{i+1}$ and $V_{i+1}$ are computed from 
$l_i, U_i$ and $V_i$ as follows:
Find any point $P$ that is at distance more than $1+\eps$ from $l_i$.
Look at the plane $h$ containing $l_i$ and $P$. We will try to set
$U_{i+1}$ and $V_{i+1}$ close to $U_h = proj(U, h)$ and $V_h = proj(V, h)$ respectively,
by the following process.

From the invariant, we have $d(U_i, U_h) \le 5$. 
So $U_h$ lies in a circle in $h$  of radius $5$ centered at $U_i$.
Create a mesh, where each element has side $\eps/8$, so that $U_i$
itself is a mesh point, and guess the mesh point closest to $U_h$ and
at a distance at most $5$ from $U_i$. We need to guess one out of
 $\frac{\pi (5)^2}{(\eps/8)^2}$ points and set this point to $U_{i+1}$.
Clearly this point is at most $\eps/4$ from $U_h$.
Similarly we guess $V_{i+1}$ out of at most $O(1/\eps^2)$ points.

\end{enumerate}

We will prove convergence by arguing that the potential function, $\Phi = d(U,U_i)^2 + d(V,V_i)^2$,
decreases significantly in each iteration. 

\begin{lemma}
We maintain the invariant, $d(U_i, U) \le 5$ and $d(V_i, V) \le 5$,  during each iteration
of algorithm $MINCYN$.
\end{lemma}
\begin{proof}
We will show that $d(U_i, U)$ only keeps decreasing and 
if the invariant is true to start with, it always remains true. Now for any point
$X$ on the plane $h$, $d(X,U)^2 = d(X,U_h)^2 + d(U_h,U)^2$. Since we choose $U_{i+1}$
to be the mesh point closest to $U_h$, among mesh points including $U_i$, $d(U_{i+1},U_h) \le d(U_i,U_h)$.
So the invariant 2 follows. 
\end{proof}

\begin{lemma}
In each iteration of algorithm $MINCYN$, the potential function $\Phi$ decreases by at least $\eps^2/2$
\end{lemma}
\begin{proof}
Note that $UU_h$ is perpendicular to the plane containing $U_h, U_i$ and $U_{i+1}$. 
So, 
  \begin{eqnarray*} 
 d(U,U_{i+1})^2 & = & d(U,U_h)^2 + d(U_h,U_{i+1})^2 \\
		& = & d(U,U_i)^2 - d(U_i,U_h)^2 + d(U_h,U_{i+1})^2 \\
		& \le & d(U,U_i)^2 - d(U_i,U_h)^2 + \eps^2/4
  \end{eqnarray*}

Similar inequality holds for $d(V,V_{i+1})^2$. Adding the two we get,
 $\Phi_{i+1} \le \Phi_{i} - d(U_i,U_h)^2 - d(V_i, V_h)^2 + \eps^2/2$.

We will prove that at least one of $d(U_i,U_h)$ and $d(V_i,V_h)$ is more than $\eps$.
For if not then we will show that $P$ cannot be within distance $1$ of any point in the
segment $U_h V_h$, which is a contradiction because $U_h V_h$ is the projection of $U V$.

Let $l_p$ be the line in plane $h$ passing through $P$ and perpendicular to $l_i$,
meeting $l_i$ at $P'$. Then, $d(P,P') \ge 1+\eps$.
Project all points to $l_p$. The segment $U_h V_h$ projects down to a segment 
$U_p V_p$. Since there is a point on $U_h V_h$ that is at most distance $1$ from $P$
there must also be such a point, $Q$, on $U_p V_p$. Since $d(P',Q) \ge \eps$,
at least one of $U_p$ and $V_p$ must be at least $\eps$ away from $P'$. Since distances
only decrease under projections, at least one of $d(U_i, U_h)$ and $d(V_i, V_h)$ must be $\ge \eps$.

So, we get $\Phi_{i+1} \le \Phi_{i} - \eps^2 + \eps^2/4 \le \Phi_{i} - \eps^2/2$ 
\end{proof}

%Finally, it is easy to show that we can chose an initial line $l_0$ and points on it $U_0$ and
%$V_0$ that satisfy the invariant - we skip the details here. 
Finally we need to prove that we can choose an initial line $l_0$ and points on it $U_0$ and
$V_0$ that satisfy the invariant. Look at any point $X$ in the set $S$. Let $Y$ be the farthest
point from $X$ in $S$. Set $l_0$ to the line passing through $X$ and $Y$. It is easy to verify
that every point in $S$ must be within distance $4$ from $l_0$ (See lemma 5.2 in \cite{HV03}
for a more general statement). Look at the projections of points in $S$ on $l_0$ and $l$.
For any point $Z$ in $S$,
let $Z_0$ denote its projection on $l_0$ and $Z_l$ denote its projection on $l$. Then
$d(Z, Z_0) \le 4$ and $d(Z, Z_l) \le 1$. So $d(Z_0, Z_l) \le 5$. 

Set $U_0$ and $V_0$ to the farthest two points
among projections of points of $S$ on $l_0$.
Clearly these points are at most at distance $5$ from $U$ and $V$ respectively.

So we have proved the following theorem
\begin{theorem}
Algorithm $MINCYN$ terminates in $O(1/\eps^2)$ iterations. That is, after so many iterations
no point will be more than $\eps$ outside the surface of the cylinder. A deterministic version of this 
algorithm  runs in time $2^{O(\frac{1}{\eps^2}\log \frac{1}{\eps})} nd$.
\end{theorem}

The deterministic version follows by simply eliminating the guess. Each guess requires guessing
twice from $O(1/\eps^2)$ choices, this guessing happens at most $O(1/\eps^2)$ times.
We also need to clarify how the optimal radius required by the algorithm can be determined.
The distance of the farthest point from the initial position of the axis $l_0$ 
is within a constant factor of the optimal radius.  The algorithm $MINCYN$ terminates
only if the radius used is larger than the optimal radius and does not terminate if
the radius used is too small.  A binary search involving
$\log(1/\eps)$ trials will result in a value that is within $1+\eps$ of the optimal radius.

Using the techniques in \cite{HV03} this algorithm generalizes to computing the min-radius
$k$-dimensional flat - we omit the details here. 
\begin{theorem}
The minimum radius $k$-dimensional flat can be computed
in time $2^{\frac{e^{O(k^2)}}{\eps^2}\log \frac{1}{\eps}} nd$.
\end{theorem}

\section*{Acknowledgments} 
I would like to thank Mihai B\u{a}doiu, Piotr Indyk, Sariel Har-Peled and Kasturi Varadharajan for
useful discussions.

\end{document}